# Equation Systems of Generalized Hydrodynamics for Soft-Matter Quasicrystals

Tian-You Fan

School of Physics, Beijing Institute of Technology, Beijing 100081, China
tyfan2013@163.com

**Abstract** The equation systems of generalized hydrodynamics, or generalized dynamics for simplicity, for soft-matter quasicrystals were established. Considering the fluidity of the matter with high order we introduced a new elementary excitation---fluid phonon to the quasicrystals, and the equation of state is necessary and introduced too. By considering other two elementary excitations---phonons and phasons, a theory---generalized hydrodynamics for soft-matter quasicrystals is set up, in which the governing equations of the dynamics are reported in this letter.

**Introduction**

In 2004，Zeng et al.[1] observed quasicrystals with 12-fold symmetry in liquid crystals. In 2005, Takano [2], in 2007, Hayashida [3] discovered a similar structure in polymers where the quasicrystals of 12-fold symmetry were also observed in calcogenides and organic dendrons. In 2009, Talapin et al. [4] found quasicrystals of 12-fold symmetry in complex binary nanoparticles.In 2011, 12- and 18-fold symmetry quasicrystals were discovered in colloids by Fischer et al. [5] who observed the structures in $PI_{30}$–$PEO_{120}$ of Poly(Isoprene-b-ethylene oxide) ($PI_n$-$PEO_m$) at room temperature by using X-ray scattering and neutron-scattering. The 18-fold symmetry quasicrystal had been observed for the first time since 1982 in solid and soft-matter quasicrystals.

The discovery of the soft-matter quasicrystals promotes the correlation between soft matter and quasicrystals, and greatly enlarged the scope of quasicrystal study. This challenges the theory and methodology of solid quasicrystal study, and provides an opportunity to develop some new theories and methods to describe matter distribution, deformation and motion of the soft-matter quasicrystals一a category of complex fluids with quasiperiodic symmetry.

To realize the task above mentioned the equation systems of the motion including an equation of state are necessary, this paper reports an attempt of ours in this direction.

It is well-known in solid quasicrystals, there are the phonon field $u_i$ and phason field $w_i$ and for the soft-matter quasicrystals, we must introduce another field, i.e., the fluid phonon field $V_i$ and their interaction among the elementary excitations, this constitutes the dynamics of the matter. The introduction of the fluid phonon changes the thermodynamics situation of the new system. In the next section we must pay attention to discuss the relevant thermodynamics of the matter.

**1. Some Information on Equation of State in Some Complex Liquids**

The soft-matter quasicrystals at first belong to soft matter which presents fluidity. The fluidity is naturally connected to fluid pressure $p$, which is related to mass density $\rho$. When we study the dynamic properties of soft-matter quasicrystals the relation between fluid pressure and mass density must be considered, this is different from that for solid quasicrystals.

The relation between the fluid pressure and mass density is the so-called equation of state $p=f(\rho)$, which is a difficult thermodynamic problem for soft matter, and beyond the scope of dynamics. In the beginning of our probe in studying the soft-matter quasicrystals we faced the difficulty, and have to take an assumption of constant relation between $p$ and $\rho$ [6], which was not a good treatment. In fact, the soft-matter quasicrystals is compressible, whose $\rho/\rho_0 \sim 10^{-3}$ which is greater 10 orders of magnitude than that of solid quasicrystals according to our systematical computation, in which $\rho_0$ denotes the initial value of the mass density. After the progressing of study on equation of state we have given the incomplete treatment up. For other complex fluid, e.g. the superfluid $^4$He Landau [7] suggested an equation of state $p = a\rho^2 + b\rho^3 + c\rho^4$, in which $a$, $b$ and $c$ are constants and determined by experiments [7]. After some decades the quantum Monte-Carlo simulation [8,9] obtained the numerical results for testing the equation that is in excellent agreement with those of experiments. These are for low temperature case. The work of the Landau school for superfluid is significant; especially the fluid phonon concept created first by his school [10] is very helpful to us in studying the soft-matter quasicrystals.

In the thermodynamic study of liquid crystals, Wensink [11] proposed an equation of state for a dense columnar liquid crystal for one-dimensional case with the normalized mass density

$$p = 3\frac{1}{\beta L}\frac{\rho}{1-\rho}, \quad \beta = 1/(k_B T)$$

in which $\rho$ is normalized, $L$ denotes the thickness of hard disk of the liquid crystals, $k_B$ the Boltzmann constant, and $T$ the absolute temperature, respectively. This equation is very simple but in the practicalapplication, there are some difficulties, which will be mentioned as below.

The author and co-worker [12] suggested an equation of state for soft-matter quasicrystals by extending the crystal results [13,14] as follows

$$p = \frac{3}{\alpha^{2/3}}\left(3\rho^{1/3} + \frac{4}{5}\frac{\beta}{\alpha}\rho^{5/6}\right) \tag{1}$$

with two constants $\alpha$ and $\beta$, which consist of 9 parameters and are quite complex and we did some relevant study[15].

Considering the complexity of equation (1), we recall the equation of Wensink but after some modifications to the original form, we obtain

$$p = 3\frac{k_B T}{l^3 \rho_0^3}\left(\rho_0^2 \rho + \rho_0 \rho^2 + \rho^3\right) \tag{2}$$

where $\rho_0$ the initial value of the mass density, $l$ the characteristic size of soft matter mentioned previously, in our computation, $l = 8 \sim 9 nm$, the theoretical prediction is the best. After the modification, the singularity of the original Wensink's equation is removed, and the appearance of possible negative pressure is also excluded. We recommendapplying the equation (2) in solving the dynamics equations.

## 2. Equations of Generalized Hydrodynamics of Soft-Matter Quasicrystals with 12-Fold Symmetry

In order to quantitatively describe the physical and mechanical properties of soft-matter quasicrystals, the governing equations of dynamics for individual systems are necessary. According to the Landau-Anderson principle of symmetry breaking and elementary excitation [16,17] and Poisson bracket of condensed matter physics[18] and the experience of Lubensky et al [19] for solid quasicrystals, the soft-matter quasicrystals of 12-fold symmetry comprise elementary excitations such as phonons, phasons and fluid phonon, and the corresponding field variables, apart from mass density $\rho$ and fluid pressure $p$, are $u_i$, $w_i$ and $V_i$, in which the latter represent fluid velocity components in the soft matter. As we know, the soft-matter quasicrystals that have been observed so far are two-dimensional quasicrystals. For two-dimensional 12-fold quasicrystals, the quasiperiodic symmetry plane is the $xy$-plane, if the $z$-axis is taken as the 12-fold symmetry axis, along the axis the atom arrangement is periodic.

Deformation and motion of soft-matter quasicrystals follow the law of mass and momentum conservations and the rule of symmetry breaking, which originated from the hydrodynamics of solid quasicrystals created by Lubensky et al. [19]. However, there are two differences between solid and soft-matter quasicrystals:

1) The solid viscosity constitutive equation in Reference [19]

$$\sigma'_{ij} = \eta_{ijkl} \dot{\xi}_{kl}, \quad \dot{\xi}_{kl} = \frac{1}{2}\left(\frac{\partial V_k}{\partial x_l} + \frac{\partial V_l}{\partial x_k}\right)$$

is replaced by the fluid constitutive equation

$$p_{ij} = -p\delta_{ij} + \sigma'_{ij} = -p\delta_{ij} + \eta_{ijkl}\dot{\xi}_{kl}, \quad \sigma'_{ij} = \eta_{ijkl}\dot{\xi}_{kl} \quad \dot{\xi}_{kl} = \frac{1}{2}\left(\frac{\partial V_k}{\partial x_l} + \frac{\partial V_l}{\partial x_k}\right);$$

so that the constitutive laws of phonons, phasons and fluid phonon are

$$\left.\begin{aligned}
\sigma_{ij} &= C_{ijkl}\varepsilon_{kl} + R_{ijkl}w_{kl}, \\
H_{ij} &= K_{ijkl}w_{ij} + R_{klij}\varepsilon_{kl}, \\
p_{ij} &= -p\delta_{ij} + \sigma'_{ij}, \sigma'_{ij} = \eta_{ijkl}\dot{\xi}_{kl}, \\
\varepsilon_{ij} &= \frac{1}{2}\left(\frac{\partial u_i}{\partial x_j} + \frac{\partial u_j}{\partial x_i}\right), w_{ij} = \frac{\partial w_i}{\partial x_j}, \dot{\xi}_{ij} = \frac{1}{2}\left(\frac{\partial V_i}{\partial x_j} + \frac{\partial V_j}{\partial x_i}\right)
\end{aligned}\right\} \quad (3)$$

respectively, where $u_i$ denotes phonon displacement vector; $\sigma_{ij}$ the phonon stress tensor; $\varepsilon_{ij}$ the phonon strain tensor; $w_i$ the phason displacement vector; $H_{ij}$ the phason stress tensor; $w_{ij}$ the phason strain tensor; $V_i$ the fluid phonon velocity vector; $p_{ij}$ the fluid stress tensor; $\rho$ the mass density; $p$ the fluid pressure; $\eta_{ijkl}$ the fluid viscosity coefficient tensor; $\dot{\xi}_{ij}$ the fluid deformation rate tensor; and $C_{ijkl}, K_{ijkl}$ and $R_{ijkl}$ are the phonon, phason and phonon-phason coupling elastic constant tensors, respectively. In addition, for 12-fold symmetry quasicrystals $R_{ijkl} = 0$ (due to decoupling between phonons and phasons). For simplicity, we only discuss the constitutive law of simplest fluid, i.e.,

$$p_{ij} = -p\delta_{ij} + \sigma'_{ij} = -p\delta_{ij} + 2\eta(\dot{\xi}_{ij} - \frac{1}{3}\dot{\xi}_{kk}\delta_{ij}) + \eta'\dot{\xi}_{kk}\delta_{ij} \qquad (4)$$

$$\sigma_{ij}^{'} = 2\eta(\dot{\xi}_{ij} - \frac{1}{3}\dot{\xi}_{kk}\delta_{ij}) + \eta^{'}\dot{\xi}_{kk}\delta_{ij} \tag{5}$$

with

$$\dot{\xi}_{kk} = \dot{\xi}_{11} + \dot{\xi}_{22} + \dot{\xi}_{33}, \quad \dot{\xi}_{ij} = \frac{1}{2}\left(\frac{\partial V_i}{\partial x_j} + \frac{\partial V_j}{\partial x_i}\right)$$

where $\eta$ is the first viscosity coefficient; $\eta^{'}$ the second one, which is omitted as it is too small(note that the description here only shows the difference of constitutive laws between solid and fluid, however this does not mean that in solid there is no pressure);

2) An equation of state $p = f(\rho)$ should be supplemented and in solid quasicrystals an equation is unnecessary. The equation of state belongs to the thermodynamics of soft matter, so the present discussion is beyond the scope of pure hydrodynamics. We used the results from Wensink [11], with some modifications by the author [12], i.e.,

$$p = f(\rho) = 3\frac{k_B T}{l^3 \rho_0^3}\left(\rho_0^2 \rho + \rho_0 \rho^2 + \rho^3\right) \tag{2}$$

where $k_B$ is the Boltzmann constant; $T$ the absolute temperature; $l$ the characteristic size of soft matter; and $\rho_0$ the initial value of mass density.

The derivation of the equations of motion is based on the Poisson bracket method of condensed matter physics [18], which was used by Lubensky et al. [19] in solid quasicrystal study for the first time. The Chinese literature on this method can be found in Reference [20] and others provided by the author, where there are many additional details on the derivations of equations of motion of quasicrystals. A key application of the Poisson bracket method lies in the Hamiltonian individual quasicrystal systems, given in the Appendix.

The equations of motion of plane field in the $xy-$ plane of soft-matter quasicrystals with 12-fold symmetry are obtained by omitting the higher terms of $\nabla_i\left(u_j \frac{\delta H}{\delta u_j}\right)$ and $\nabla_i\left(w_j \frac{\delta H}{\delta w_j}\right)$ listed in the Appendix, and the derivation details in mathematics are not included:

$$\left.\begin{aligned}
&\frac{\partial \rho}{\partial t}+\nabla \cdot(\rho \mathbf{V})=0 \\
&\frac{\partial(\rho V_x)}{\partial t}+\frac{\partial(V_x \rho V_x)}{\partial x}+\frac{\partial(V_y \rho V_x)}{\partial y}=-\frac{\partial p}{\partial x}+\eta \nabla^2 V_x+\frac{1}{3}\eta\frac{\partial}{\partial x}\nabla \cdot \mathbf{V}+M\nabla^2 u_x+(L+M-B)\frac{\partial}{\partial x}\nabla \cdot \mathbf{u} \\
&-(A-B)\frac{1}{\rho_0}\frac{\partial \delta\rho}{\partial x} \\
&\frac{\partial(\rho V_y)}{\partial t}+\frac{\partial(V_x \rho V_y)}{\partial x}+\frac{\partial(V_y \rho V_y)}{\partial y}=-\frac{\partial p}{\partial y}+\eta \nabla^2 V_y+\frac{1}{3}\eta\frac{\partial}{\partial y}\nabla \cdot \mathbf{V}+M\nabla^2 u_y+(L+M-B)\frac{\partial}{\partial y}\nabla \cdot \mathbf{u} \\
&-(A-B)\frac{1}{\rho_0}\frac{\partial \delta\rho}{\partial y} \\
&\frac{\partial u_x}{\partial t}+V_x\frac{\partial u_x}{\partial x}+V_y\frac{\partial u_x}{\partial y}=V_x+\Gamma_{\mathbf{u}}\left[M\nabla^2 u_x+(L+M)\frac{\partial}{\partial x}\nabla \cdot \mathbf{u}\right] \\
&\frac{\partial u_y}{\partial t}+V_x\frac{\partial u_y}{\partial x}+V_y\frac{\partial u_y}{\partial y}=V_y+\Gamma_{\mathbf{u}}\left[M\nabla^2 u_y+(L+M)\frac{\partial}{\partial y}\nabla \cdot \mathbf{u}\right] \\
&\frac{\partial w_x}{\partial t}+V_x\frac{\partial w_x}{\partial x}+V_y\frac{\partial w_x}{\partial y}=\Gamma_{\mathbf{w}}\left[K_1\nabla^2 w_x+(K_2+K_3)\frac{\partial}{\partial y}\left(\frac{\partial w_x}{\partial y}+\frac{\partial w_y}{\partial x}\right)\right] \\
&\frac{\partial w_y}{\partial t}+V_x\frac{\partial w_y}{\partial x}+V_y\frac{\partial w_y}{\partial y}=\Gamma_{\mathbf{w}}\left[K_1\nabla^2 w_y+(K_2+K_3)\frac{\partial}{\partial x}\left(\frac{\partial w_x}{\partial y}+\frac{\partial w_y}{\partial x}\right)\right] \\
&p=f(\rho)=3\frac{k_B T}{l^3 \rho_0^3}\left(\rho_0^2 \rho+\rho_0 \rho^2+\rho^3\right)
\end{aligned}\right\} \quad (6)$$

where $\nabla^2=\frac{\partial^2}{\partial x^2}+\frac{\partial^2}{\partial y^2}$, $\nabla=\mathbf{i}\frac{\partial}{\partial x}+\mathbf{j}\frac{\partial}{\partial y}$, $\mathbf{V}=\mathbf{i}V_x+\mathbf{j}V_y$, $\mathbf{u}=\mathbf{i}u_x+\mathbf{j}u_y$, and $L=C_{12}$, $M=(C_{11}-C_{12})/2$ are the phonon elastic constants; $K_1, K_2, K_3$ the phason elastic constants; $\eta$ the fluid dynamic viscosity; $\Gamma_u$ and $\Gamma_w$ the phonon and phason dissipation coefficients; $A$ and $B$ the material constants due to variation of mass density; $k_B$ the Boltzmann constant; $T$ the absolute temperature; $l$ the characteristic size of soft matter; and $\rho_0$ the initial value of the mass density.

Equations (6) is the final governing equations in the plane field of dynamics of soft-matter quasicrystals of 12-fold symmetry with field variables $u_x, u_y, w_x, w_y, V_x, V_y, \rho$ and $p$. The number of the field variables is eight, as is the amount of field equations: (6a) is the mass conservation equation or continuum equation; (6b) and (6c) are momentum conservation equations or generalized Navier-Stokes equations; (6d) and (6e) are the equations of motion of phonons due to symmetry breaking; (6f) and (6g) are phason dissipation equations; and (6h) is the equation of state. These equations reveal the nature of wave propagation of fields $\mathbf{u}$ and $\mathbf{v}$, and the nature of diffusion of field $\mathbf{w}$ from the view of hydrodynamics. The equations are consistent with mathematical solvability. If there is lack of the equation of state, the equation system is not closed, and has no meaning mathematically and physically. This shows that the equation of state is necessary. However, the derivation of equations of motion of above given (see Appendix) does not include the equation of state (which belongs to thermodynamics). The original source of equation of state is from Wensink [11], and Fan and Fan [12] and later receiveda modification to obtain the present version of the equation.

## 3. Equations of Generalized Hydrodynamics of Soft-Matter Quasicrystals with 18-Fold Symmetry

In solid quasicrystals, there has yet to be any 18-fold symmetry observed. However, Hu et al. [21] predicted the existence of the structure, and put forward a six-dimensional embedded space theory to describe its phonon and phason excitations, especially they pointed out there are two types of phasons, i.e., the first and second phason fields $v_i$ and $w_i$ respectively. Fan accepted the theory of Hu et al and promoted the application of the theory in his new monographs [22], Li and Fan [23] developed the theoryof dislocations of the quasicrystals. The author studied the theory of soft-matter quasicrystals with 18-fold symmetry, in which there are phonon field $u_i$, first and second phason fields $v_i$ and $w_i$, in addition to a fluid phonon field $V_i$ for the soft-matter quasicrystals, respectively. According to the new theory there are constitutive laws for 18-fold symmetry soft-matter quasicrystals:

$$\left.\begin{aligned}
\sigma_{ij} &= \frac{\partial F}{\partial \varepsilon_{ij}} = C_{ijkl}\varepsilon_{kl} + r_{ijkl}v_{kl} + R_{ijkl}w_{kl} \\
\tau_{ij} &= \frac{\partial F}{\partial v_{ij}} = T_{ijkl}v_{kl} + r_{klij}\varepsilon_{kl} + G_{ijkl}w_{kl} \\
H_{ij} &= \frac{\partial F}{\partial w_{ij}} = K_{ijkl}w_{kl} + R_{klij}\varepsilon_{kl} + G_{klij}v_{kl} \\
p_{ij} &= -p\delta_{ij} + \sigma_{ij}', \sigma_{ij}' = \eta_{ijkl}\dot{\xi}_{kl} \\
\varepsilon_{ij} &= \frac{1}{2}\left(\frac{\partial u_i}{\partial x_j} + \frac{\partial u_j}{\partial x_i}\right), v_{ij} = \frac{\partial v_i}{\partial x_j}, w_{ij} = \frac{\partial w_i}{\partial x_j}, \dot{\xi}_{ij} = \frac{1}{2}\left(\frac{\partial V_i}{\partial x_j} + \frac{\partial V_j}{\partial x_i}\right)
\end{aligned}\right\} \quad (7)$$

where $u_i$ denotes phonon displacement vector; $\sigma_{ij}$ the phonon stress tensor; $\varepsilon_{ij}$ the phonon strain tensor; $v_i$ the first phason displacement vector; $\tau_{ij}$ the first phason stress tensor; $v_{ij} = \frac{\partial v_i}{\partial x_j}$ the first phason strain tensor; $w_i$ the second phason displacement vector; $H_{ij}$ the associate phason stress tensor; $w_{ij} = \frac{\partial w_i}{\partial x_j}$ the second phason strain tensor; $V_i$ the fluid phonon velocity vector; $p_{ij}$ the fluid stress tensor; $p$ the fluid pressure; $\eta_{ijkl}$ the fluid viscosity coefficient tensor; $\dot{\xi}_{ij}$ the fluid deformation rate tensor; $C_{ijkl}, T_{ijkl}, K_{ijkl}$ the phonon, first and second phason elastic constant tensors, respectively; $r_{ijkl}, R_{ijkl}, G_{ijkl}$ the elastic constant tensors of phonon-first phason coupling, phonon-second phason coupling, and first-second phason coupling, respectively. Furthermore, for 18-fold symmetry quasicrystals, $r_{ijkl} = R_{ijkl} = 0$ due to the decoupling between phonons and phasons. For simplicity, we only discuss the simplest fluid, i.e.,

$$p_{ij} = -p\delta_{ij} + \sigma_{ij}' = -p\delta_{ij} + 2\eta(\dot{\xi}_{ij} - \frac{1}{3}\dot{\xi}_{kk}\delta_{ij}) + \eta'\dot{\xi}_{kk}\delta_{ij},$$

$$\dot{\xi}_{kk} = \dot{\xi}_{11} + \dot{\xi}_{22} + \dot{\xi}_{33}, \quad \dot{\xi}_{ij} = \frac{1}{2}\left(\frac{\partial V_i}{\partial x_j} + \frac{\partial V_j}{\partial x_i}\right)$$

where $\eta$ is the first viscosity coefficient, and $\eta'$ is the second one, which was omitted, as it is too small. By extending the derivation adopted in the previous sections, we obtained the equations of dynamics of soft-matter quasicrystals with 18-fold symmetry as follows:

$$\begin{aligned}
&\frac{\partial \rho}{\partial t} + \nabla \cdot (\rho \mathbf{V}) = 0 \\
&\frac{\partial(\rho V_x)}{\partial t} + \frac{\partial(V_x \rho V_x)}{\partial x} + \frac{\partial(V_y \rho V_x)}{\partial y} = -\frac{\partial p}{\partial x} + \eta \nabla^2 V_x + \frac{1}{3}\eta \frac{\partial}{\partial x}\nabla \cdot \mathbf{V} + M\nabla^2 u_x + (L+M-B)\frac{\partial}{\partial x}\nabla \cdot \mathbf{u} \\
&\quad -(A-B)\frac{1}{\rho_0}\frac{\partial \delta \rho}{\partial x} \\
&\frac{\partial(\rho V_y)}{\partial t} + \frac{\partial(V_x \rho V_y)}{\partial x} + \frac{\partial(V_y \rho V_y)}{\partial y} = -\frac{\partial p}{\partial y} + \eta \nabla^2 V_y + +\frac{1}{3}\eta \frac{\partial}{\partial y}\nabla \cdot \mathbf{V} + M\nabla^2 u_y + (L+M-B)\frac{\partial}{\partial y}\nabla \cdot \mathbf{u} \\
&\quad -(A-B)\frac{1}{\rho_0}\frac{\partial \delta \rho}{\partial y} \\
&\frac{\partial u_x}{\partial t} + V_x\frac{\partial u_x}{\partial x} + V_y\frac{\partial u_x}{\partial y} = V_x + \Gamma_\mathbf{u}\left[M\nabla^2 u_x + (L+M)\frac{\partial}{\partial x}\nabla \cdot \mathbf{u}\right] \\
&\frac{\partial u_y}{\partial t} + V_x\frac{\partial u_y}{\partial x} + V_y\frac{\partial u_y}{\partial y} = V_y + \Gamma_\mathbf{u}\left[M\nabla^2 u_y + (L+M)\frac{\partial}{\partial y}\nabla \cdot \mathbf{u}\right] \\
&\frac{\partial v_x}{\partial t} + V_x\frac{\partial v_x}{\partial x} + V_y\frac{\partial v_x}{\partial y} = \Gamma_v\left[T_1\nabla^2 v_x + G\left(\frac{\partial^2 w_x}{\partial x^2} - \frac{\partial^2 w_x}{\partial y^2}\right) - 2G\frac{\partial^2 w_y}{\partial x \partial y}\right] \\
&\frac{\partial v_y}{\partial t} + V_x\frac{\partial v_y}{\partial x} + V_y\frac{\partial v_y}{\partial y} = \Gamma_v\left[T_1\nabla^2 v_y + 2G\frac{\partial^2 w_x}{\partial x \partial y} + G\left(\frac{\partial^2 w_y}{\partial x^2} - \frac{\partial^2 w_y}{\partial y^2}\right)\right] \\
&\frac{\partial w_x}{\partial t} + V_x\frac{\partial w_x}{\partial x} + V_y\frac{\partial w_x}{\partial y} = \Gamma_\mathbf{w}\left[K_1\nabla^2 w_x + G\left(\frac{\partial^2 v_x}{\partial x^2} - \frac{\partial^2 v_x}{\partial y^2}\right) + 2G\frac{\partial^2 v_y}{\partial x \partial y}\right] \\
&\frac{\partial w_y}{\partial t} + V_x\frac{\partial w_y}{\partial x} + V_y\frac{\partial w_y}{\partial y} = \Gamma_\mathbf{w}\left[K_1\nabla^2 w_y - 2G\frac{\partial^2 v_x}{\partial x \partial y} + G\left(\frac{\partial^2 v_y}{\partial x^2} - \frac{\partial^2 v_y}{\partial y^2}\right)\right] \\
&p = f(\rho) = 3\frac{k_B T}{l^3 \rho_0^3}\left(\rho_0^2 \rho + \rho_0 \rho^2 + \rho^3\right)
\end{aligned} \quad (8)$$

Equation set (8) is the final governing equation system of generalized dynamics of soft-matter quasicrystals with 18-fold symmetry with field variables $u_x, u_y, v_x, v_y, w_x, w_y, V_x, V_y, \rho$ and $p$. Both number of field variables and field equations was 10. Among the field equations: (8a) is the mass conservation equation or the continuum equation; (8b) and (8c) are momentum conservation equations or generalized Navier-Stokes equations; (8d) and (8e) are equations of motion of phonons due to symmetry breaking; (8f) and (8g) are first phason dissipation equations; and (8h) and (8i) are the second phason dissipation equations; and (8j) is the equation of state. These equations reveal the nature of wave propagation of fields $\mathbf{u}$ and $\mathbf{V}$, and the nature of diffusion of fields $\mathbf{v}$ and $\mathbf{w}$ from the view of hydrodynamics. The equations are consistent with mathematical solvability: If there is a lack of the equation of state, the equation system is not closed, and has no meaning mathematically and physically. This demonstrates that the equation of state is necessary.

## 4. Equations of Generalized Hydrodynamics of Possible Soft-Matter Quasicrystals with 5-and 10-Fold Symmetries

The 12- and 18-fold symmetrical soft-matter quasicrystals were observed so far, with a possibility that 5- and 10-fold symmetrical soft-matter quasicrystals will be discovered thereafter. These two kinds of solid quasicrystals are very stable, which promotesimportant interest. Especially as there are strong coupling effects between the phonons and phasons, it is interesting to study their mechanical and physical properties and mathematical solutions. If we consider the plane field in the $xy-$plane to be a quasiperiodic plane, and if $z-$axis is the 5- or 10-fold symmetry axis, then after derivation similar to those previous sections, we obtain the final governing equation system for these two kinds of soft-matter quasicrystals:

$$\left.\begin{aligned}
&\frac{\partial \rho}{\partial t}+\nabla\bullet(\rho\mathbf{V})=0\\
&\frac{\partial(\rho V_x)}{\partial t}+\frac{\partial(V_x\rho V_x)}{\partial x}+\frac{\partial(V_y\rho V_x)}{\partial y}=-\frac{\partial p}{\partial x}+\eta\nabla^2 V_x+\frac{1}{3}\eta\frac{\partial}{\partial x}\nabla\bullet\mathbf{V}+M\nabla^2 u_x+(L+M-B)\frac{\partial}{\partial x}\nabla\bullet\mathbf{u}\\
&+R_1\left(\frac{\partial^2 w_x}{\partial x^2}+2\frac{\partial^2 w_y}{\partial x\partial y}-\frac{\partial^2 w_x}{\partial y^2}\right)-R_2\left(\frac{\partial^2 w_y}{\partial x^2}-2\frac{\partial^2 w_x}{\partial x\partial y}-\frac{\partial^2 w_y}{\partial y^2}\right)-(A-B)\frac{1}{\rho_0}\frac{\partial\delta\rho}{\partial x}\\
&\frac{\partial(\rho V_y)}{\partial t}+\frac{\partial(V_x\rho V_y)}{\partial x}+\frac{\partial(V_y\rho V_y)}{\partial y}=-\frac{\partial p}{\partial y}+\eta\nabla^2 V_y+\frac{1}{3}\eta\frac{\partial}{\partial y}\nabla\bullet\mathbf{V}+M\nabla^2 u_y+(L+M-B)\frac{\partial}{\partial y}\nabla\bullet\mathbf{u}\\
&+R_1\left(\frac{\partial^2 w_y}{\partial x^2}-2\frac{\partial^2 w_x}{\partial x\partial y}-\frac{\partial^2 w_y}{\partial y^2}\right)+R_2\left(\frac{\partial^2 w_x}{\partial x^2}+2\frac{\partial^2 w_y}{\partial x\partial y}-\frac{\partial^2 w_x}{\partial y^2}\right)-(A-B)\frac{1}{\rho_0}\frac{\partial\delta\rho}{\partial y}\\
&\frac{\partial u_x}{\partial t}+V_x\frac{\partial u_x}{\partial x}+V_y\frac{\partial u_x}{\partial y}=V_x+\Gamma_\mathbf{u}[M\nabla^2 u_x+(L+M)\frac{\partial}{\partial x}\nabla\bullet\mathbf{u}+\\
&R_1\left(\frac{\partial^2 w_x}{\partial x^2}+2\frac{\partial^2 w_y}{\partial x\partial y}-\frac{\partial w_x}{\partial y^2}\right)-R_2\left(\frac{\partial^2 w_y}{\partial x^2}-2\frac{\partial^2 w_x}{\partial x\partial y}-\frac{\partial^2 w_y}{\partial y^2}\right)]\\
&\frac{\partial u_y}{\partial t}+V_x\frac{\partial u_y}{\partial x}+V_y\frac{\partial u_y}{\partial y}=V_y+\Gamma_\mathbf{u}[M\nabla^2 u_y+(L+M)\frac{\partial}{\partial y}\nabla\bullet\mathbf{u}+\\
&R_1\left(\frac{\partial^2 w_y}{\partial x^2}-2\frac{\partial^2 w_x}{\partial x\partial y}-\frac{\partial^2 w_y}{\partial y^2}\right)+R_2\left(\frac{\partial^2 w_x}{\partial x^2}+2\frac{\partial^2 w_y}{\partial x\partial y}-\frac{\partial^2 w_x}{\partial y^2}\right)]\\
&\frac{\partial w_x}{\partial t}+V_x\frac{\partial w_x}{\partial x}+V_y\frac{\partial w_x}{\partial y}=\Gamma_\mathbf{w}[K_1\nabla^2 w_x+\\
&R_1\left(\frac{\partial^2 u_x}{\partial x^2}-2\frac{\partial^2 u_y}{\partial x\partial y}-\frac{\partial^2 u_x}{\partial y^2}\right)+R_2\left(\frac{\partial^2 u_y}{\partial x^2}+2\frac{\partial^2 u_x}{\partial x\partial y}-\frac{\partial^2 u_y}{\partial y^2}\right)]\\
&\frac{\partial w_y}{\partial t}+V_x\frac{\partial w_y}{\partial x}+V_y\frac{\partial w_y}{\partial y}=\Gamma_\mathbf{w}[K_1\nabla^2 w_y+\\
&R_1\left(\frac{\partial^2 u_y}{\partial x^2}+2\frac{\partial^2 u_x}{\partial x\partial y}-\frac{\partial^2 u_y}{\partial y^2}\right)-R_2\left(\frac{\partial^2 u_x}{\partial x^2}-2\frac{\partial^2 u_y}{\partial x\partial y}-\frac{\partial^2 u_x}{\partial y^2}\right)]\\
&p=f(\rho)=3\frac{k_B T}{l^3\rho_0^3}\left(\rho_0^2\rho+\rho_0\rho^2+\rho^3\right)
\end{aligned}\right\}(9)$$

These equations reveals the nature of wave propagation of fields $\mathbf{u}$ and $\mathbf{V}$, and the nature of the diffusion of field $\mathbf{w}$ from the view of hydrodynamics.

## 5. Equations of Generalized Hydrodynamics of Possible Soft-Matter Quasicrystals with8-Fold Symmetry

Apart from the observed 12- and 18-fold symmetrical soft-matter quasicrystals, and potential5- and 10-fold symmetrical soft-matter quasicrystals yet tobe discovered, the 8-fold symmetrical soft-matter quasicrystals may also be observed in thenear future. This kind of solid quasicrystal is very stable, which is importantespecially as there are strong coupling effects between the phonons and phasons, and it is interesting to study their mechanical and physical properties and mathematical solutions. We considered the plane of quasiperiodicity to be $xy-$plane, if the $z-$axis is 8-fold symmetry axis. Next, for the possibility of soft-matter octagonal quasicrystals in soft matter there is the final governing equation system of the generalized dynamics, after some derivations by the Poisson bracket method of condensed matter physics:

$$\left.\begin{aligned}
&\frac{\partial \rho}{\partial t}+\nabla \cdot(\rho \mathbf{V})=0 \\
&\frac{\partial(\rho V_x)}{\partial t}+\frac{\partial(V_x \rho V_x)}{\partial x}+\frac{\partial(V_y \rho V_x)}{\partial y}=-\frac{\partial p}{\partial x}+\eta\nabla^2 V_x+\frac{1}{3}\eta\frac{\partial}{\partial x}\nabla\cdot\mathbf{V}+M\nabla^2 u_x+(L+M-B)\frac{\partial}{\partial x}\nabla\cdot\mathbf{u} \\
&\qquad\qquad +R\left(\frac{\partial^2 w_x}{\partial x^2}+2\frac{\partial^2 w_y}{\partial x\partial y}-\frac{\partial^2 w_x}{\partial y^2}\right)-((A-B)\frac{1}{\rho_0}\frac{\partial\delta\rho}{\partial x} \\
&\frac{\partial(\rho V_y)}{\partial t}+\frac{\partial(V_x \rho V_y)}{\partial x}+\frac{\partial(V_y \rho V_y)}{\partial y}=-\frac{\partial p}{\partial y}+\eta\nabla^2 V_y+\frac{1}{3}\eta\frac{\partial}{\partial y}\nabla\cdot\mathbf{V}+M\nabla^2 u_y+(L+M-B)\frac{\partial}{\partial y}\nabla\cdot\mathbf{u} \\
&\qquad\qquad +R\left(\frac{\partial^2 w_y}{\partial x^2}-2\frac{\partial^2 w_x}{\partial x\partial y}-\frac{\partial^2 w_y}{\partial y^2}\right)-(A-B)\frac{1}{\rho_0}\frac{\partial\delta\rho}{\partial y} \\
&\frac{\partial u_x}{\partial t}+V_x\frac{\partial u_x}{\partial x}+V_y\frac{\partial u_x}{\partial y}=V_x+\Gamma_{\mathbf{u}}[M\nabla^2 u_x+(L+M)\frac{\partial}{\partial x}\nabla\cdot\mathbf{u}+R\left(\frac{\partial^2 w_x}{\partial x^2}+2\frac{\partial^2 w_y}{\partial x\partial y}-\frac{\partial w_x}{\partial y^2}\right)] \\
&\frac{\partial u_y}{\partial t}+V_x\frac{\partial u_y}{\partial x}+V_y\frac{\partial u_y}{\partial y}=V_y+\Gamma_{\mathbf{u}}[M\nabla^2 u_y+(L+M)\frac{\partial}{\partial y}\nabla\cdot\mathbf{u}+R_1\left(\frac{\partial^2 w_y}{\partial x^2}-2\frac{\partial^2 w_x}{\partial x\partial y}-\frac{\partial^2 w_y}{\partial y^2}\right)] \\
&\frac{\partial w_x}{\partial t}+V_x\frac{\partial w_x}{\partial x}+V_y\frac{\partial w_x}{\partial y}=\Gamma_{\mathbf{w}}[K_1\nabla^2 w_x+(K_2+K_3)\left(\frac{\partial^2 w_x}{\partial y^2}+\frac{\partial^2 w_y}{\partial x\partial y}\right)+R\left(\frac{\partial^2 u_x}{\partial x^2}-2\frac{\partial^2 u_y}{\partial x\partial y}-\frac{\partial^2 u_x}{\partial y^2}\right)] \\
&\frac{\partial w_y}{\partial t}+V_x\frac{\partial w_y}{\partial x}+V_y\frac{\partial w_y}{\partial y}=\Gamma_{\mathbf{w}}[K_1\nabla^2 w_y+(K_2+K_3)\left(\frac{\partial^2 w_x}{\partial x\partial y}+\frac{\partial^2 w_y}{\partial x^2}\right)+R\left(\frac{\partial^2 u_y}{\partial x^2}+2\frac{\partial^2 u_x}{\partial x\partial y}-\frac{\partial^2 u_y}{\partial y^2}\right)] \\
&p=f(\rho)=3\frac{k_B T}{l^3\rho_0^3}\left(\rho_0^2\rho+\rho_0\rho^2+\rho^3\right)
\end{aligned}\right\} \quad (10)$$

Again, these equations reveal the nature of wave propagation of fields **u** and **V**, and the nature of diffusion of field **w** from the view of hydrodynamics.

## 6. Conclusion and Discussion

The governing equations (6) for dynamics of soft-matter quasicrystals of 12-fold symmetry and (8) for soft-matter quasicrystals of 18-fold symmetry are given, these two types structures and materials are observed in liquids, polymers, colloids, nanoparticles and surfactants etc., the governing equations (9) and (10) for dynamics of possible soft-matter quasicrystals of 5-/ 10-fold and 8-fold symmetry, which may be observed in the near future.

The Equation systems (6),(8), and (9), (10) are the governing equation systems of generalized hydrodynamics of soft-matter quasicrystals observed and possibly discovered. These are novel nonlinear partial differential equations andprovide thebasis for studying dynamics of soft-matter quasicrystals.

The aim ofsetting up these equations lies in describing the matter distribution, deformation and motion of these new materials, and to do so, we must solve initial- and boundary-value problems of these equations. Computation shows these equations to beconsistent and mathematicallysolvable.

This work opens a new research direction for science and engineering of soft-matter quasicrystals serving basic study and technological applications.

As the solutions require the assistance of mathematical physics and computational physics, and need a large volume for presentation, many details have not included due to space limitations.

**Acknowledgements** The author thanks the National Natural Science Foundation of China for the support through Grant 11272053; thanks also to Professors T C Lubensky at the University of Pennsylvania, USA; Stephen Z D Cheng at the University of Akron, USA; H H Wensink at the Utrecht University in The Netherlands; and Xian-Fang Li at the Central South University in China for beneficial discussions and kind assistance.

**Appendix The differential-variational form of equations of motion of hydrodynamics of soft-matter quasicrystals (it does not include the equation of state)**

The equations of motion of soft-matter quasicrystals are derived by the Poisson bracket method, which is the heritage and development of Lubensky et al for solid quasicrystals [19], in which the key is the Hamiltonians for individual quasicrystal systems. For the first kind of two-dimensional quasicrystals of soft matter, the energy functional or the Hamiltonians are similar to that given by Lubensky et al for solid quasicrystals in form for the first kind of two-dimensional quasicrystals in soft matter such as:

$$H = H[\Psi(\mathbf{r},t)] = \int \frac{\mathbf{g}^2}{2\rho} d^d \mathbf{r} + \int \left[ \frac{1}{2} A \left( \frac{\delta\rho}{\rho_0} \right)^2 + B \left( \frac{\delta\rho}{\rho_0} \right) \nabla \cdot \mathbf{u} \right] d^d \mathbf{r} + F_{el}$$

$$= H_{kin} + H_{density} + F_{el} \tag{A1}$$

$$\mathbf{g} = \rho \mathbf{V}, \quad F_{el} = F_u + F_w + F_{uw}$$

where $F_{el}$ denotes the elastic strain energy; and $F_u, F_w, F_{uw}$ represent the strain energies of phonons, phasons and phonon-phason coupling of the matter for 5-, 8-,10- and 12-fold symmetry quasicrystals, respectively:

$$F_u = \int \frac{1}{2} C_{ijkl} \varepsilon_{ij} \varepsilon_{kl} d^d \mathbf{r}$$

$$F_w = \int \frac{1}{2} K_{ijkl} w_{ij} w_{kl} d^d \mathbf{r} \tag{A2}$$

$$F_{uw} = \int \left( R_{ijkl} \varepsilon_{ij} w_{kl} + R_{klij} w_{ij} \varepsilon_{kl} \right) d^d \mathbf{r}$$

Due to the difference of constitutive laws between soft-matter quasicrystals and solid quasicrystals, the results of the Hamiltonians for soft-matter quasicrystals are different from those of solid quasicrystals.

For the 18-fold symmetry quasicrystals of soft matter, the elastic energy will be

$$F_{el} = F_u + F_v + F_w + F_{uv} + F_{uw} + F_{vw} \tag{A3}$$

where $F_u, F_v, F_w, F_{uv}, F_{uw}, F_{vw}$ represent the strain energies of phonons, first phasons, second phasons, phonon-first phason coupling, phonon-second phason coupling, and first phason-second phason coupling, respectively:

$$F_u = \int \frac{1}{2} C_{ijkl} \varepsilon_{ij} \varepsilon_{kl} d^d \mathbf{r}$$

$$F_v = \int \frac{1}{2} T_{ijkl} v_{ij} v_{kl} d^d \mathbf{r}$$

$$F_w = \int \frac{1}{2} K_{ijkl} w_{ij} w_{kl} d^d \mathbf{r} \quad (A4)$$

$$F_{uv} = \int \left( r_{ijkl} \varepsilon_{ij} v_{kl} + r_{klij} v_{ij} \varepsilon_{kl} \right) d^d \mathbf{r}$$

$$F_{uw} = \int \left( R_{ijkl} \varepsilon_{ij} w_{kl} + R_{klij} w_{ij} \varepsilon_{kl} \right) d^d \mathbf{r}$$

$$F_{vw} = \int \left( G_{ijkl} v_{ij} w_{kl} + G_{klij} w_{ij} v_{kl} \right) d^d \mathbf{r}$$

Therefore, the problem of the second kind of quasicrystals is more complex than that of the first kind.

With the Hamiltonian (by using the Poisson bracket method) and after lengthy derivation we have the equations of motion for the second kind of two-dimensional quasicrystals in soft matter as follows

$$\frac{\partial \rho}{\partial t} + \nabla_k (\rho V_k) = 0$$

$$\frac{\partial g_i(\mathbf{r},t)}{\partial t} = -\nabla_k (\mathbf{r})(V_k g_i) + \nabla_j (\mathbf{r})\left(-p\delta_{ij} + \eta_{ijkl} \nabla_k(\mathbf{r}) V_l \right) - \left(\delta_{ij} - \nabla_i u_j \right) \frac{\delta H}{\delta u_j(\mathbf{r},t)}$$

$$+ \left( \nabla_i v_j \right) \frac{\delta H}{\delta v_j(\mathbf{r},t)} + \left( \nabla_i w_j \right) \frac{\delta H}{\delta w_j(\mathbf{r},t)} - \rho \nabla_i (\mathbf{r}) \frac{\delta H}{\delta \rho(\mathbf{r},t)}, \qquad g_j = \rho V_j$$

$$\frac{\partial u_i(\mathbf{r},t)}{\partial t} = -V_j \nabla_j(\mathbf{r}) u_i - \Gamma_u \frac{\delta H}{\delta u_i(\mathbf{r},t)} + V_i$$

$$\frac{\partial v_i(\mathbf{r},t)}{\partial t} = -V_j \nabla_j(\mathbf{r}) v_i - \Gamma_v \frac{\delta H}{\delta v_i(\mathbf{r},t)} \quad (A5)$$

$$\frac{\partial w_i(\mathbf{r},t)}{\partial t} = -V_j \nabla_j(\mathbf{r}) w_i - \Gamma_w \frac{\delta H}{\delta w_i(\mathbf{r},t)}$$

The equations for the first kind of two-dimensional quasicrystals in soft matter can be obtained by omitting the field variable $v_i$. Of course, these equations of motion do not include the equation of state, which is a thermodynamic result rather than the result of derivation based on the Poisson bracket. The terms of variational form can be reduced to a differential form based on Reference [14] given by Lubensky.

After further simplifications, from Equation (A5) and including Equation (5), the equation of state, we can obtain the generalized dynamics Equations (6), (8), (9) and (10) for the plane field of individual two-dimensional case.

**After the publication of this paper the author and his group have obtained many solutions of some initial and boundary value problems of equations (6), (8), (9) and (10), which examine the equations, the examination shows the equations are correct and effective, please refer to the new publications of the author and his co-workers in the field.**